# Clarification of size effects in polycrystalline BaTiO$_3$ thin films by means of the specific heat measurements: grain size or film thickness?

**B.A. Strukov, S.T. Davitadze, S.G. Shulman*, B.M. Goltzman* and V.V. Lemanov***

*Lomonosov Moscow State University, Moscow 119992, Russia (bstrukov@mail.ru)*
**Ioffe Physico-Technical Institute, St.-Petersburg 194021, Russia*

Specific heat of polycrystalline BaTiO$_3$ thin films on the fused quartz substrate was measured by ac-hot probe method. Phase transition temperature, excess entropy and spontaneous polarization were determined as a function of film thickness and grain size. The variation of the latter was obtained in the limits 30 – 150 nm by changing of the temperature of the substrate during sputtering while thickness of films 20 – 1100 nm was controlled by the conditions of sputtering. It was found that the relation between the thickness and grain size is important for the size effects in polycrystalline films.

## INTRODUCTION

The peculiarities of phase transitions in submicron ferroelectric films are of increasing interest both for theory and applications. There is a variety of experimental data concerning ferroelectric films, which doesn't allow one to have the definite conclusions about finite size effects even for the monocrystalline epitaxial films. It is hard to realize the perfect control of the technology conditions of sputtering of films; film-substrate misfit strains which in principle supposed to affect order and temperature of the phase transition [1-3], might be partly or completely weakened by a domain or dislocation structure formation [4,5] and can be considered as a "free parameter" of the problem. In the theoretical estimations it is not quite clear if the widely used phenomenological description of the size effect [6-8] can be applied to the nanoscale objects although it qualitatively predicts the set of effects which can be supported by experiment. It is an even more complicate situation with the polycrystalline thin films where one has two characteristic sizes – crystallite size and thickness of the film. The phenomenological approach was applied to analyze the specific case of the free standing ferroelectric film with the cell structure in the form of a rectangular periodical pattern in the film plane [9]. Therefore the "doubled" size effect was predicted resulting from both film thickness and the lateral column size.

It should be noted that up to now there are quite a few experimental data in favor or against the results of the phenomenological theory mentioned above. The goal of this work was to clear up at least partly the situation. Specific heat measurements were used as the tool for the revealing of evolution of ferroelectric phase transition in $BaTiO_3$ polycrystalline thin films with a variation of the film thickness and grain size.

## EXPERIMENTAL

$BaTiO_3$ films were deposited on fused quartz substrate (typical size $4.0 \times 5.0 \times 0.5$ mm$^3$) by rf magnetron sputtering (f = 13.6 MHz) with the substrate position "off axis" with different substrate temperature. X-ray diffraction revealed {111} textured polycrystalline film with a tetragonal symmetry. A mixture of $Ar:O_2$ in a ratio 0.8:0.2 was introduced and the total pressure was held at $4 \times 10^{-2}$ Torr.

Two series of samples were obtained with a variation of the sputtering conditions – temperature of substrate and power of magnetron. In the first series the films of different thickness were obtained at T = 750 $^0$C; deposition rate was estimated by the control of sputtering time as 5 nm/min. The films with the thickness in the range 20 – 1100 nm were obtained. AMF control allowed us to reveal the colunmar structure of the films with an average column diameter about 150 nm for the films with thickness 70 – 1100 nm.

In the second series we changed both the temperature of deposition and magnetron power. Keeping constant thickness of the film (about 500 nm) we were able to change the column diameter in the range 30 – 150 nm that was controlled by analysis of an AMF images of films.

The experimental setup was described in details in [10]. The ac-hot probe (3$\omega$) method was used for determination of the temperature dependence of specific heat of $BaTiO_3$ deposited on a fused quartz substrate. Because of the fact that the thermal contrast between films and substrate was large enough, it was possible to measure the important thermodynamic parameters of the films – the location of $T_C$, absolute values of the specific heat in the region of ferroelectric phase transition, the excess entropy of the transition.

## EXPERIMENTAL RESULTS AND DISCUSSION

The temperature dependence of $BaTiO_3$ specific heat for the films with different thickness is shown in Fig. 1. It is clear that when the thickness of the film reduces, the phase transition temperature decreases while a diffusion of the anomaly increases. The anomaly is quite weak for the 40 nm film; we have not detected any anomaly for 20 nm film. It should be noted that sharp increase of a film roughness was revealed for the ul­trathin films [10].

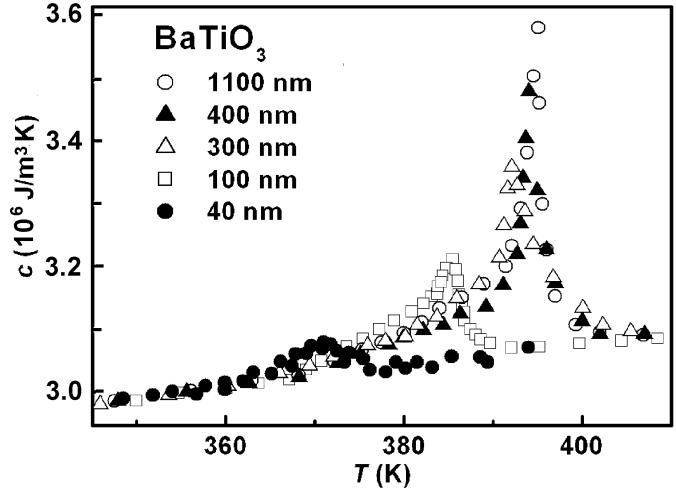

**Fig. 1.** Temperature depend­ence of the specific heat of BaTiO$_3$ films with different thickness near ferroelectric phase transition point.

The anomalies of the specific heat can be used to determine transition temperature and excess entropy of transitions as functions of the film thickness. Fig.2 illustrates the de­pendence of the temperature and excess entropy of the ferroelectric phase transition in BaTiO$_3$ films on their thickness (T$_C$ was determined as the temperature of the maximum value of the specific heat); it is evident that the spontaneous polarization of the films which is connected with the specific heat by the simple thermodynamic relation is pro­gressively suppressed as the thickness is reduced.

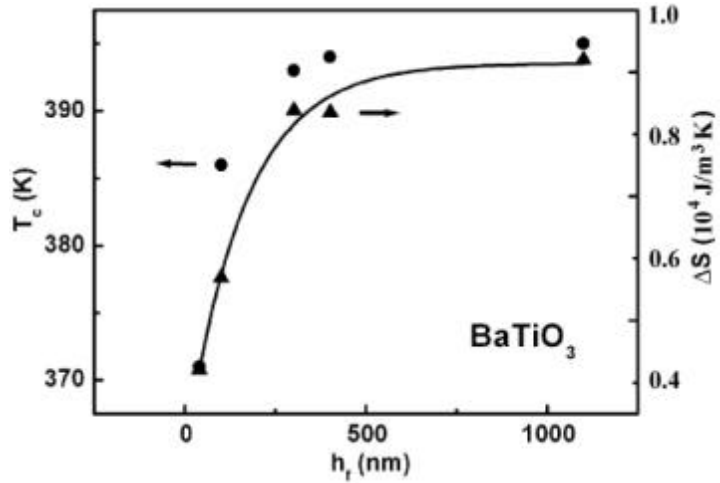

**Fig. 2.** Temperature and ex­cess entropy of the ferroelec­tric phase transition in BaTiO$_3$ films as a function of thick­ness.

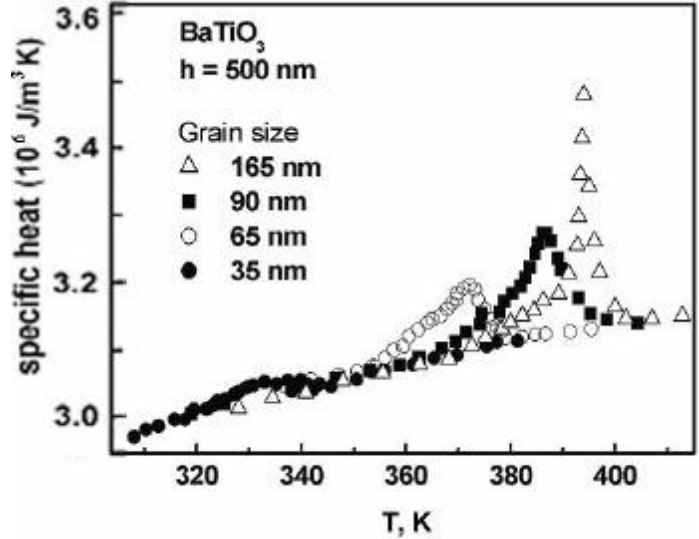

**Fig. 3.** Temperature dependence of the specific heat of 500 nm BaTiO$_3$ films with the different grain size.

Fig. 3 shows the effect of the grain size variation upon the anomaly of the specific heat for the same film thickness (approximately 500 nm). In this series the film thickness can be considered as constant and "thick" enough and the size effect is definitely connected only with the grain pattern scale. The sharp decrease of the transition temperature, excess entropy as well as suppressing of the spontaneous polarization can be detected in this case and illustrated by Fig. 4.

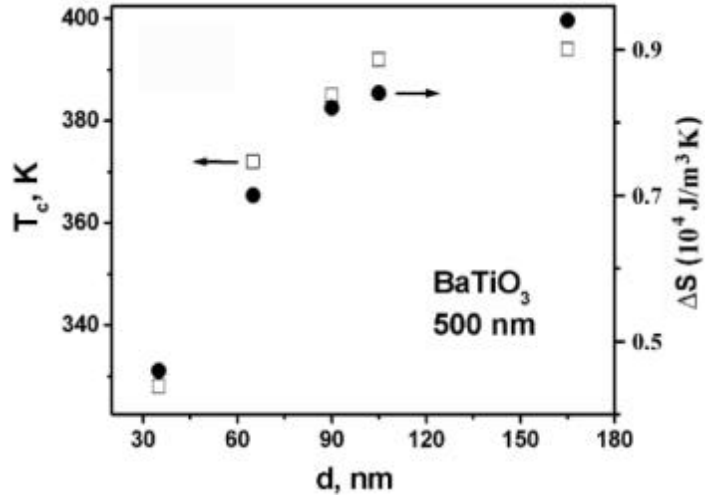

**Fig. 4.** Temperature and excess entropy of the ferroelectric phase transition in 500 nm BaTiO$_3$ films as a function of the grain size.

It is interesting to note that the transition temperature is found to be the linear function of both reciprocal film thickness and grain size (Fig.5). Therefore for the both cases we can reveal the critical size for a room temperature size-driven phase transition (10 nm for thickness and 32 nm for grain size) and for the complete suppressing of ferroelectric activity (2.5 nm for thickness and 8 nm for grain size).

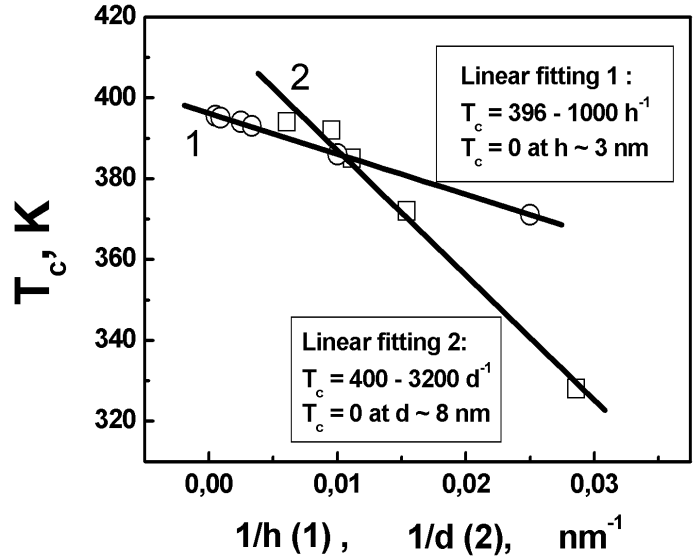

**Fig. 5.** Dependence of the ferroelectric phase transition temperature on the reciprocal film thickness (1) and grain size (2) for BaTiO$_3$ films.

In conclusion we have shown by means of dynamical ac-hot probe calorimetry that the size-driven effects in polycrystalline BaTiO$_3$ films are equally connected with two parameters – a film thickness and grain size. To obtain a comprehensive picture of the phase transition in such films one have to obtain the full three-dimensional presentation of the transition point as a function of the film thickness and grain size. This work is now in progress.

**ACKNOWLEDGMENTS**

The authors are grateful to Russian Foundation for Basic Research (Projects 03-02-17518 and 03-02-06841)